\documentclass[prb,twocolumn,floats,aps]{revtex4}
\usepackage{graphicx,graphics,color,epsfig} 
\usepackage{bm}
\usepackage{amssymb}
\usepackage{amsmath}
\usepackage{amsfonts}

\newcommand{\mi}{{\,\mathrm{i}}}

\def\frac#1#2{{#1\over#2}}

\begin{document}

\title{Effect of nonlocal interactions on the disorder-induced zero-bias anomaly in the Anderson-Hubbard model}
\author{Hong-Yi Chen}
\affiliation{Department of Physics, National Taiwan Normal University, Taipei 11677, Taiwan}
\author{W.A.\ Atkinson, and R.\ Wortis}
\affiliation{Department of Physics \& Astronomy, Trent University,
1600 West Bank Dr., Peterborough ON, K9J 7B8, Canada}
\date{\today}

\begin{abstract}
To expand the framework available for interpreting experiments on disordered strongly correlated systems, and in particular to explore further the strong-coupling zero-bias anomaly found in the Anderson-Hubbard model, we ask how this anomaly responds to the addition of nonlocal electron-electron interactions.  
We use exact diagonalization to calculate the single-particle density of states of the extended Anderson-Hubbard model.
We find that for weak nonlocal interactions the form of the zero-bias anomaly is qualitatively unchanged.
The energy scale of the anomaly continues to be set by an effective hopping amplitude renormalized by the nonlocal interaction.
At larger values of the nonlocal interaction strength, however, hopping ceases to be a relevant energy scale and higher energy features associated with charge correlations dominate the density of states.
\end{abstract}
\maketitle

\section{Introduction}
\label{sec-intro}

As the simplest model which incorporates strong correlations, the Hubbard model is widely used as a starting point for exploring the diverse behaviors of transition metal oxides.  
Understanding the effect of disorder in these systems is of interest given the importance of chemical doping in tuning their properties.
In fact, nanoscale electronic disorder has been observed and explicitly correlated with the locations of dopant atoms in some materials.\cite{Schmidt2011} 
Disorder can be incorporated into the Hubbard model, and the most widely-studied version is the Anderson-Hubbard model (AHM), which is defined below.
The combination of local interactions and disorder in the Anderson-Hubbard model has been shown to result in a suppression of the density of states (DOS) at the Fermi level, a feature known as a zero-bias anomaly (ZBA).
The ZBA in the AHM has a number of unique features:
The DOS suppression occurs in the absence of nonlocal interactions, in contrast with other, well-known examples including both the Efros-Shklovskii Coulomb gap\cite{Efros1975} and the Altshuler-Aharonov anomaly\cite{Altshuler1985}.  
Moreover, the energy scale in the AHM ZBA is proportional to the hopping amplitude $t$ and independent of both the local interaction $U$ and the disorder strength $\Delta$ within a broad range of phase space.

The strong-coupling ZBA appears to arise from kinetic energy savings, rather than a reduction of the Coulomb energy.\cite{Wortis2010}  In the absence of hopping, electronic states are confined to individual atomic sites.  With nonzero $t$, these states can extend over nearby sites, lowering the electronic kinetic energy.  In the large $U$ limit, however, this spreading of electronic wave functions is strongly inhibited by the local Coulomb interaction. Nonetheless, if the disorder is sufficiently strong, there will be atomic sites for which the energy of double occupancy is nearly degenerate with the energy to form a singlet with one of its neighbors. For these special configurations, the local Coulomb interaction does not inhibit electronic motion between the two atoms, and the energy of the system is reduced by an amount of order $t$ relative to the atomic case.\cite{Wortis2010,Chen2010}  This leads directly to the suppression of spectral weight at the Fermi energy over an energy scale $t$.  This mechanism is unique to strongly-correlated systems.

Given the presence of nonlocal interactions in real materials, as well as their importance in theories of ZBAs in other models, it is natural to ask how nonlocal interactions influence the kinetic-energy driven ZBA in the AHM.
This formal question bears on a number of classes of materials.
(1) Most directly this work relates to doped transition metal oxides.
DOS measurements\cite{Sarma1998,Maiti2007} in SrRu$_{1-x}$Ti$_x$O$_3$ and LaNi$_{1-x}$Mn$_x$O$_3$ show deviations from the standard pictures of Efros-Shklovskii\cite{Efros1975} and Altshuler-Aharonov\cite{Altshuler1985}, and it is an open question whether this is because of strong correlation physics.
In addition, early work on the ZBA in the AHM suggested that disorder may contribute to the stability of the pseudogap in high temperature superconductors.\cite{Chiesa2008}
(2) A second class of materials to which this work may be relevant are dilute doped semiconductors and granular metals.
Generally described using atomic-limit models, these systems display Coulomb gap behavior.
Open questions include whether there is an association between the Coulomb gap and glassy behavior,\cite{Pankov2005,Surer2009,Goethe2009} and how electron mobility (and hence screening) influence the observed behaviors.\cite{Goethe2009,Delahaye2010}  
(3) A third class of materials are two-dimensional electron gas systems, such as thin metal films and MOSFETs.  Whereas (noninteracting) localization theory concludes that there should be no metal-insulator transition in two dimensions, experiments on these systems suggest otherwise.\cite{Kravchenko2004}  
Recently it has been proposed that the insulating behavior in these films is in fact due not to disorder but to interactions.\cite{Amaricci2010}
To support this proposal, the extended Hubbard model was explored, but only in the clean limit.
Even if interactions drive the insulating behavior, the systems remain disordered.  
(4) Finally, organic conductors are another class of materials in which strong correlations can be important.
Very recently it has been shown that disorder can be introduced into organic conductors by x-ray irradiation, resulting in novel behaviors and expanded opportunities for exploring interactions and disorder.\cite{Sasaki2008, Sano2010}

\begin{figure}[htbp] 
   \centering
   \includegraphics[width=\columnwidth]{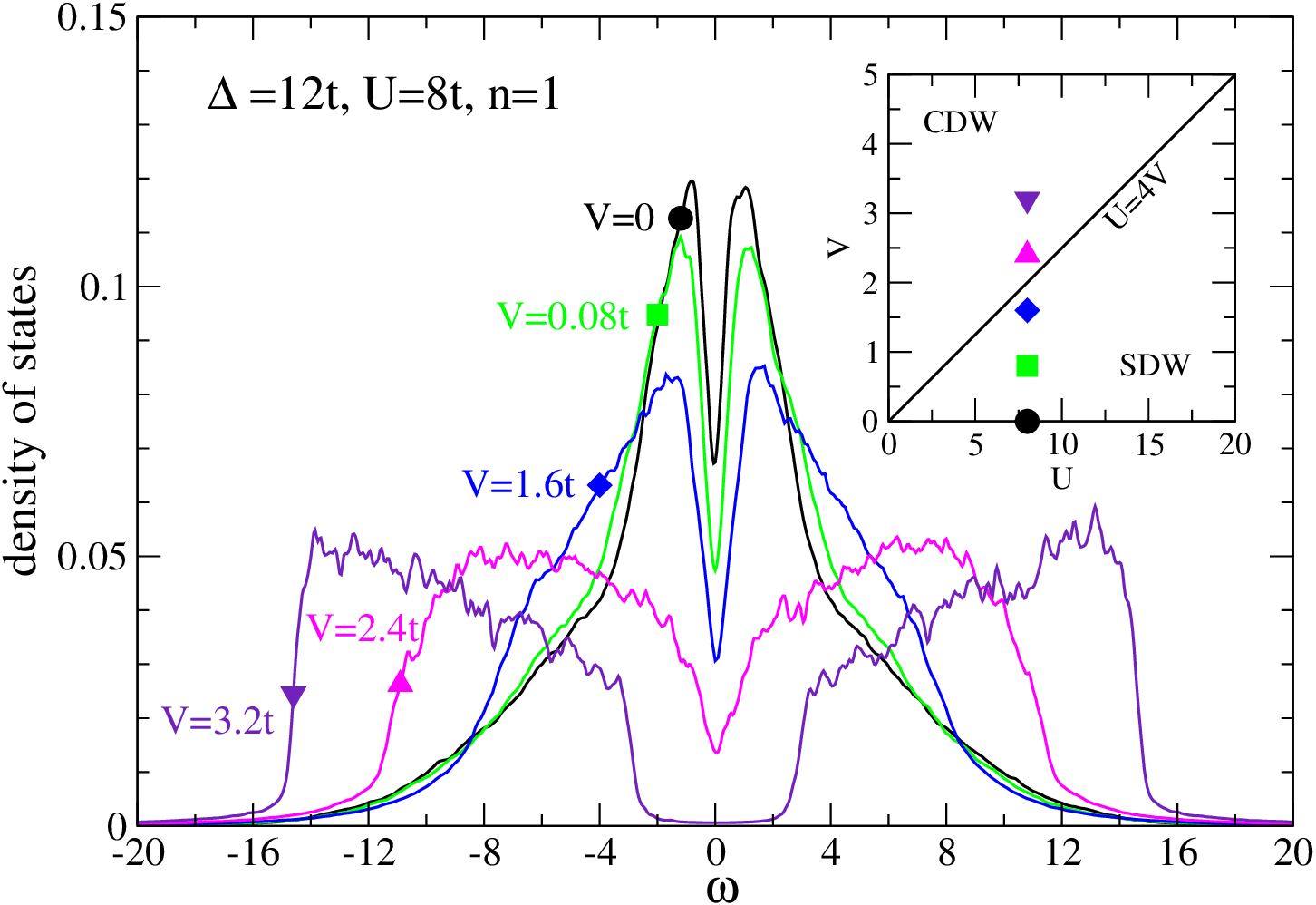} 
   \caption{Density of states versus frequency at half filling ($n=1$) for $\Delta=12t$, $U=8t$ and $V$ values as indicated.  Throughout this work, results are for 12-site lattices averaged over 1000 disorder configurations.  Inset indicates the location of each parameter set on the phase diagram of the {\em clean} extended Anderson-Hubbard model.}
   \label{half_fill_Vdep}
\end{figure}

To expand the theoretical framework available for interpreting these diverse materials, we explore the DOS of the extended Anderson-Hubbard model (EAHM) on a number of trajectories in the available phase space, all in the strong disorder limit.  
Figure \ref{half_fill_Vdep} summarizes our main result, namely the presence of a crossover in the form of the DOS as a function of interaction strength.  
When the nonlocal interactions are weak, electron mobility plays a key role, generating in particular the narrow kinetic-energy-driven ZBA seen earlier in the AHM.\cite{Chiesa2008,Wortis2010}
In the half-filled case shown in Fig.\ \ref{half_fill_Vdep}, when nonlocal interactions are strong, atomic-limit physics dominates the DOS:  A broad suppression around the Fermi level is associated with charge correlations driven by the nearest-neighbor repulsion.
Doping away from half filling reduces the impact of interactions and hence attenuates this effect.
%

Section \ref{sec-meth} describes our approach while our results are presented and discussed in Section \ref{sec-randd}.

\section{Method}
\label{sec-meth}

The extended Anderson-Hubbard model includes the hopping $t$ and onsite Coulomb repulsion $U$ of the traditional Hubbard model as well as two additional terms:  a nearest-neighbor Coulomb repulsion $V$ and disordered site potentials $\epsilon_i$.
\begin{eqnarray}
\mathcal{H}
&=& 
-t \sum_{\langle i,j \rangle,\sigma} {\hat c}_{i\sigma}^{\dag} {\hat c}_{j\sigma}
+ \sum_\mi U {\hat n}_{i\uparrow} {\hat n}_{i\downarrow} \nonumber \\
& & 
+ \sum_{\langle i,j \rangle} {V \over 2} {\hat n}_i {\hat n}_j 
+ \sum_{i,\sigma} \epsilon_i {\hat n}_{i\sigma},
\end{eqnarray}
We consider a two-dimensional square lattice.
${\hat c}_{i\sigma}^{\dag}$ is the creation operator for lattice site $i$ and spin $\sigma$.
${\hat n}_{i\sigma}={\hat c}_{i\sigma}^{\dag} {\hat c}_{i\sigma}$.  
$\langle i,j \rangle$ refers to nearest neighbor pairs.
The site potentials $\epsilon_i$ are chosen from a flat distribution of width $\Delta$: $P(\epsilon_i)=\Theta(\Delta/2 - |\epsilon_i|)/\Delta$ where $\Theta$ is the Heaviside function.  
We focus on the limit of strong disorder:  All results shown are for $\Delta=12$ in units of the hopping $t$.

The Lanczos method is used to calculate the DOS of 12-site clusters.
The Lanczos method\cite{Dagotto1994} denotes a collection of iterative procedures all founded on the idea that a matrix $\underline{\underline Q}$ can be found such that 
$\underline{\underline Q}^{\dag} \ \underline{\underline H} \ \underline{\underline Q} = \underline{\underline T}$
where $\underline{\underline T}$ is a tridiagonal matrix.  
Computational savings come from the fact that the number of columns $n$ in $\underline{\underline Q}$ may be less than the number of rows such that $\underline{\underline T}$ is smaller than $\underline{\underline H}$.  
The extremal eigenvalues of $\underline{\underline T}$ converge quickly to those of $\underline{\underline H}$ as a function of increasing $n$.  
Finding the DOS using the Lanczos method proceeds in two steps.
First, restarted Lanczos is used to find the ground state $|\Psi_0\rangle$ and energy $E_0$. 
Second block-recursion\cite{Golub1996} is used to calculate the Green's function.

The LDOS at site $i$ of a particular disorder configuration $c$ is given by
\begin{eqnarray}
\rho_{ci}(\omega) &=& - {1 \over \pi} {\rm Im} \ G_{ii}^c(\omega).
\end{eqnarray}
where 
\begin{eqnarray}
G_{ii}^c(\omega) &=& \langle \psi_0^c | {\hat c}_i [\omega +E_0^c- {\hat H} + i \eta]^{-1} {\hat c}_i^{\dag} | \psi_0^c \rangle
\nonumber \\
& &
+ \langle \psi_0^c | {\hat c}_i^{\dag} [\omega -E_0^c+ {\hat H} + i \eta]^{-1} {\hat c}_i | \psi_0^c \rangle
\end{eqnarray}
is the $i^{\rm th}$ diagonal element of the real-space Green's function.
Here $ |\psi_o^c \rangle $ and $E_0^c$ are the ground state wave function and the ground state energy of disorder configuration $c$.
The DOS of a single disorder configuration is 
\begin{eqnarray}
\rho_c(\omega) &=& {1 \over N_s} \sum_{i=1}^{N_s} \rho_{ci}(\omega)
\end{eqnarray}
where $N_s=12$ is the number of sites in the lattice.
We present DOS results averaged over many disorder configurations
\begin{eqnarray}
\rho(\omega) &=& {1 \over N_{config}} \sum_{c=1}^{N_{config}} \rho_c(\omega)
\end{eqnarray}
The number of disorder configurations $N_{config}=1000$ for all results presented here.
Because we study systems with very strong disorder, the mean free path is of order the lattice spacing, and hence the disorder averaged DOS can be expected to be representative of the DOS in the thermodynamic limit.  Comparison of Lanczos results on a 10-site cluster with determinant quantum Monte Carlo results on a 64-site system by Chiesa, et al\cite{Chiesa2008} support this.  

\section{Results and Discussion}
\label{sec-randd}

To explore how nonlocal interactions affect the ZBA in the Anderson-Hubbard model, we present here DOS results first for the case of half filling, followed by lower dopings.

\subsection{Half filling}

\begin{figure}[htbp] 
   \centering
   \includegraphics[width=\columnwidth]{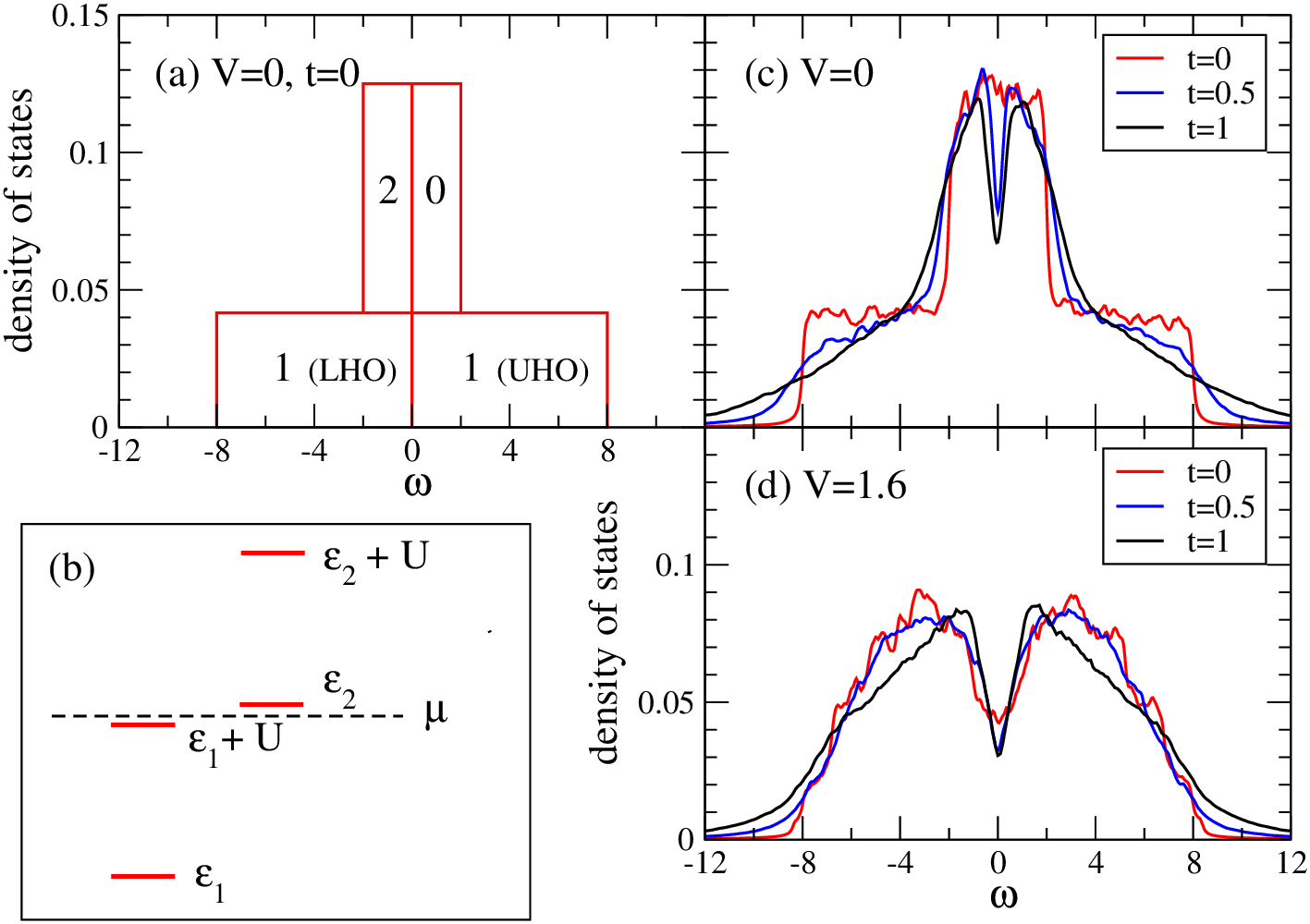} 
   \caption{DOS versus frequency at half filling ($n=1$) for $\Delta=12$ and $U=8$.  
   (a) Analytic result for $V=0$ and $t=0$ with the occupancies of contributing sites indicated.
   (b) Diagram showing an arrangement of atomic orbitals for which the introduction of hopping moves DOS weight away from the Fermi level.
   (c) Numerical results for $V=0$ and $t$ values as indicated.
   (d) Numerical results for $V=1.6$ and $t$ values as indicated.}
   \label{tdep}
\end{figure}

\subsubsection{Review of the $V=0$ case}

Before proceeding to the new results, let us review what is known about the $V=0$ case.
The highest curve in Fig.\ \ref{half_fill_Vdep} shows the ZBA found in the AHM, which was explored by Chiesa, et al.\cite{Chiesa2008}
For sufficiently strong disorder, the width of this ZBA is independent of disorder strength, interaction strength, and doping.  
The one parameter which controls the width is the hopping $t$, on which the width depends linearly.  
The existence of a ZBA in the absence of nonlocal interactions is novel.
The two standard frameworks for discussing ZBAs were developed by Efros and Shklovskii\cite{Efros1975} and by Altshuler and Aronov.\cite{Altshuler1985}
Efros and Shklovskii\cite{Efros1975} addressed a system with $1/r$ Coulomb interactions and disordered site potentials in
the atomic limit.
They argued that the DOS at an energy $\epsilon$ is proportional to $|\epsilon-\epsilon_F|^{d-1}$ where $\epsilon_F$ is the Fermi level and $d$ is the dimension.
The zero DOS at the Fermi level is known as a soft gap or more specifically as the Coulomb gap, 
and its existence depends on the infinite range of the Coulomb interaction.
Altshuler and Aronov\cite{Altshuler1985} considered the limit of weak disorder and weak interactions, and used diagrammatic perturbation theory to show that in this limit a cusp appears in the DOS near the Fermi level. 
The result is not especially sensitive to the form of the interaction.
However, for purely local interactions the correction to the DOS is positive;
DOS suppression only occurs when nonlocal interactions are present.
In both the Altshuler-Aronov and Efros-Shklovskii pictures nonlocal interactions are key to the suppression of the DOS at the Fermi level, yet in the AHM there is a ZBA.
Moreover, although a dependence on the effective nonlocal interaction $J \propto t^2/U$ might be expected, here the dominant energy scale is $t$ alone.

To understand this linear dependence on $t$, it is useful to start from the atomic limit and then consider what happens as hopping is turned on.\cite{Wortis2010,Chen2010,Wortis2011,Chen2011}  
In the atomic limit, each site contributes to the DOS at, at most, two energies:  the site potential $\epsilon_i$, and $\epsilon_i+U$.  
It is convenient to refer to these as the lower Hubbard orbital and the upper Hubbard orbital.
When the site is singly occupied, it contributes to the DOS at both energies because a particle may be either added or removed.  
When the site is empty (doubly occupied) only the addition (removal) of a particle is possible and hence the DOS contribution is only at $\epsilon_i$ ($\epsilon_i+U$).
In the ground state, sites with potentials $\epsilon_i>\mu$ are empty, those for which $\mu>\epsilon_i>\mu-U$ are singly occupied, and those for which $\mu-U>\epsilon_i$ are doubly occupied.
Putting these together, the atomic-limit DOS corresponding to any combination of $\Delta$, $U$ and $\mu$ may be constructed.
The case of $\Delta=12$, $U=8$ and $\mu=4$ (half filling) is shown in Fig.\ \ref{tdep}(a).
The numbers in each block indicate the ground state occupancy of the sites which contributed.
An important point for our purposes is that there is no ZBA at zero temperature in the atomic limit.
Nonzero temperature does suppress the DOS even in the atomic limit,\cite{Wortis2011} but we restrict ourselves here to zero temperature.

We now ask how hopping affects the DOS.  
To address this it is convenient to consider the simple case of a two-site system.
Consider in particular the configuration shown in Fig.\ \ref{tdep}(b):  $\epsilon_1+U$ is just below the chemical potential, and $\epsilon_2$ is just above.
In the atomic limit, the first site will be doubly occupied and the second empty, corresponding to the Fock state $|20 \rangle$ and to a grand potential $E-\mu N$ near zero.
However, when hopping is allowed, the new ground state will be a linear combination of the singlet states
$(|\uparrow \downarrow\rangle - |\downarrow \uparrow \rangle)/\sqrt{2}$ and $|2 0 \rangle$.
The grand potential of this ground state is lower than the atomic one by $t$,
because the probability amplitude for finding an electron is now spread over both sites lowering the kinetic energy of the many-body state.
To linear order in $t$, there is no corresponding shift in the 1-particle and 3-particle excited states.  
The energy of transitions is therefore increased, corresponding to a shift of the poles in the Green's function away from the Fermi level.
If we consider an ensemble of such two-site systems, the result is the opening of a ZBA in the DOS of width $t$.\cite{Wortis2010,Chen2010}

While larger lattices present many additional complications,\cite{Chen2011} this simple two-site picture demonstrates how kinetic-energy savings can lead to a ZBA of width $t$.
It is worth emphasizing that this behavior is unique to strongly correlated systems because it relies on there being a large difference in energy between the atomic lower and upper Hubbard orbitals at each site.  
For this reason, the effect is not captured by mean-field treatments.
Finally, we note that the effect requires double occupancy on some nonzero fraction of sites.
Fig.\ \ref{tdep}(a) shows that, in the atomic limit when both empty and doubly occupied sites are present, the Fermi level falls somewhere within the high central-plateau.
When the chemical potential is lowered such that no sites are doubly occupied, the Fermi level instead sits at the left edge of this central plateau.
When hopping is nonzero but still much less than $U$ and $\Delta$, this feature of the central plateau in the DOS persists, and the position of the Fermi level relative to this central plateau continues to be indicative of the level of double occupancy.
Because of the importance of double occupancy in the formation of the ZBA, the linear $t$ dependence of the ZBA is only expected when the Fermi level falls within this central plateau.\cite{Wortis2010}

\subsubsection{Evolution of the DOS with $V$}

Having reviewed the $V=0$ case, we return to Fig.\ \ref{half_fill_Vdep}.
The remaining curves demonstrate the evolution of the DOS as a function of the nearest-neighbor interaction strength $V$.  
Qualitatively, the lower-$V$ curves resemble the $V=0$ curve and are distinct from those at higher $V$ values.
For $V=0.8$ and $V=1.6$ the most obvious changes are in the height and width of the central peak, while the form of the ZBA is relatively consistent.
For $V=2.4$ and $V=3.2$, however, there is a strong shift of spectral weight away from the Fermi level, such that the DOS is largest near the band edges.
Moreover, the ZBA loses its sharp form and opens into a hard gap.
The sections below provide a more detailed discussion first of the distinct physics present when $V \ne 0$, then of the small $V$ behavior and finally of the large $V$ behavior.

\subsubsection{Comparing $V=0$ with $V \ne 0$} 

A striking feature of Fig.\ \ref{half_fill_Vdep}, as noted above, is the similarity in the shape of the ZBAs seen at $V=0.8$ and $1.6$ with that at $V=0$.
Fig.\ \ref{tdep} emphasizes an important distinction between the case of $V=0$ and that of $V \ne 0$.

For the case $V=0$, Fig.\ \ref{tdep}(c) shows the DOS with and without hopping.
The $t=0$ curve shows the wedding-cake structure predicted in Fig.\ \ref{tdep}(a), with no ZBA.
When $t$ is turned on, a ZBA emerges with a width linear in $t$ as described above.

In contrast, when $V \ne 0$, Fig.\ \ref{tdep}(d) shows that there is a ZBA even in the atomic limit.
This rounded anomaly at $t=0$ is a manifestation of the same atomic limit physics found in the Efros-Shklovskii Coulomb gap, but for a short-range interaction.
The DOS is not suppressed to zero here because the interaction range is finite.

When hopping is turned on, the shape of the ZBA changes abruptly.
Note that the $t=0.5$ curve coincides with the $t=1$ curve at low energies and with the $t=0$ curve at higher energies.  
This suggests that the reshaping of the ZBA by hopping begins at the Fermi level and spreads outward in energy as $t$ is increased.
The atomic limit ($t=0$) is classical in the sense that only integer occupancy is allowed.
It appears that the quantum effects introduced by hopping have their first effect on the DOS at the Fermi level, while classical behavior persists at higher energies.

In summary, Fig.\ \ref{tdep} emphasizes the sharp distinction in the atomic limit between having nonlocal interaction and not.
The addition of hopping generates a new ZBA which is qualitatively similar with and without nonlocal interactions.

\subsubsection{Small $V$}
\label{small_V}

\begin{figure}[htbp] 
   \centering
   \includegraphics[width=\columnwidth]{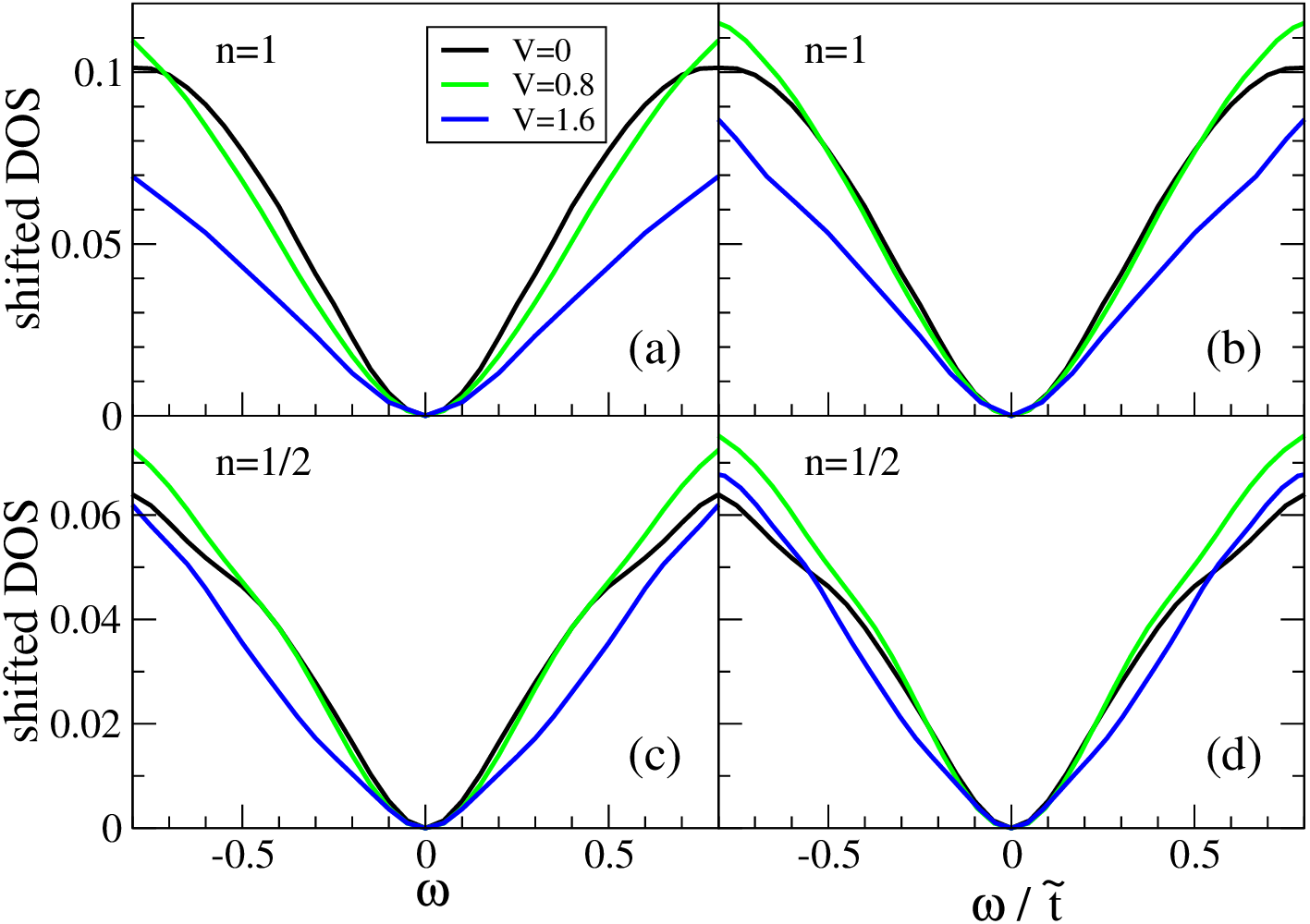} 
   \caption{Symmetrized DOS versus frequency in units of $t$ at half filling (a) and at quarter filling (c) shifted to coincide at the Fermi level.  $\Delta=12t$, $U=8t$, and $V$ values as indicated.
Panels (b) and (d) show the same data plotted versus frequency in units of the mean-field rescaled hopping $\tilde t$.}
   \label{rescale}
\end{figure}

We now turn to a more quantitative comparison of the three qualitatively similar curves: $V=0,\ 0.8$ and 1.6.
Fig.\ \ref{rescale} (a) shows the three curves vertically shifted to coincide at the Fermi level.
Here we see that the ZBA is becoming broader as $V$ is increased.
For these small $V$ values, a mean-field treatment of the nonlocal interaction provides some insight.
A mean-field treatment of the nearest-neighbor interaction results in the following expression:
\begin{eqnarray}
{V \over 2} \sum_{\langle i j \rangle} {\hat n}_i {\hat n}_j 
&  \approx& 
V \sum_{\langle i j \rangle} 
\biggl[
n_j {\hat n}_i
- \sum_{\sigma} f_{ji} {\hat c}_{i\sigma}^{\dag} {\hat c}_{j\sigma} \nonumber \\
& & \hskip 0.5 in 
- {n_i n_j \over 2} + f_{ij} f_{ji} \biggr]
\end{eqnarray}
where $n_i \equiv \sum_{\sigma} \langle {\hat c}_{i\sigma}^{\dag} {\hat c}_{i\sigma} \rangle$
and $f_{ji\sigma} \equiv \langle {\hat c}_{j\sigma}^{\dag} {\hat c}_{i\sigma} \rangle$
with $\langle ... \rangle$ denoting the expectation value with respect to the ground state.
The last two terms are constants which simply shift the zero of energy.
The first term in this expression renormalizes the site energies $\epsilon_i$ consistent with the idea that interactions screen the disorder potential.
However, when $V$ is very small relative to the disorder potential and $U$ is large such that the charge density is very uniform, this term will have very little effect.
Our focus here is on the second term which results in a renormalized hopping integral:
\begin{eqnarray*}
t \rightarrow {\tilde t} = t + V f_{ji}
\label{renormt}
\end{eqnarray*}
where $i$ and $j$ are nearest neighbor sites.
This is consistent with the gradual increase in the width of the anomaly as $V$ is increased.

More precisely, Fig.\ \ref{rescale}(b) shows the $V=0$, 0.8 and 1.6 curves with the frequency axis in units of $\tilde t$.  
Under this rescaling, the $V=0.8$ curve coincides very closely with the $V=0$ curve in the frequency range of the ZBA. 
The $V=1.6$ curve also matches but only at very low energies.  
That this mean-field approach works less well at $V=1.6$ than at $V=0.8$ is due in part to the other changes $V$ causes in the system as discussed further in the next section.  
However, another issue here is that in calculating $\tilde{t}$ for this figure we used $f_{ij}$ averaged over all bonds in the lattice.  With more computational effort, a value for $\tilde{t}$ more specifically associated with the sites which contribute to the DOS near $\omega=0$ could be constructed.  We expect $f_{ij}$ to be larger on bonds between sites with orbital configurations as in Fig.\ \ref{tdep}(b), and hence we expect that this improved mean-field treatment would create a stronger rescaling resulting in a better match at $V=1.6$.

\subsubsection{Large $V$}

Interaction strength generally falls off with distance, making $V$ values near $U$ unphysical.  However, well before this cutoff, qualitatively distinct behavior arises:
The $V=2.4$ and 3.2 curves in Fig.\ \ref{half_fill_Vdep} are very different from those at lower $V$ values.  There is a strong shift of spectral weight well away from the Fermi level such that the DOS is largest near the band edges.

A useful point of reference is the clean extended Hubbard model.  In two dimensions, this model has a first-order phase transition between charge density wave (CDW) order for $U<4V$ and spin density wave (SDW) order for $U>4V$.\cite{Aichhorn2004}
Fig.\ \ref{half_fill_Vdep} inset shows this clean phase diagram with the locations corresponding to the DOS curves marked.

How does disorder affect this phase diagram?
Quenched disorder influences first-order phase transitions in a wide variety of ways depending on the details of the model.\cite{Imry1979}
One possibility is that the phase boundary may be moved.  However, in our case the boundary is between two ordered phases, neither of which is enhanced by the disorder.
Indeed, the fact that the abrupt change in the shape of our DOS occurs between $V=1.6$ and $V=2.4$ suggests that $U=4V$ remains significant.
The bigger issue is whether regions of order remain on either side of this line.

On the $U=0$ axis, a connection may be made with the well studied random-field Ising model (RFIM).  
The atomic limit of the EAHM at $U=0$ is ${\cal H} = \sum_i \epsilon_i n_i + V \sum_{\langle i,j\rangle} n_i n_j$.
Using $n_i=S_i+1$, this becomes ${\cal H} = \sum_i \epsilon_i S_i + V \sum_{\langle i,j \rangle} S_i S_j +$ constants (for fixed particle number).  
When $U=0$, site occupancies $n_i$ are restricted to 0 and 2 at zero temperature.  
Therefore, this is precisely the RFIM with $\epsilon_i$ playing the role of the local field and $V$ the spin interaction favoring antiferromagnetic (AFM) order.  
In two dimensions, the RFIM is always disordered.\cite{Natterman1998} 
Although a more rigorous proof has been developed\cite{Aizenman1989}, this is most easily illustrated by the following surface-to-volume argument.\cite{Natterman1998} 
Imagine the system begins with perfect AFM order, and then consider flipping all the spins in a domain of size $L^d$ where $d$ is the dimension.  Such a flip will raise the energy associated with the interaction term by an amount proportional to the length of the boundary: $L^{d-1}$.  This flip will also change the energy associated with the random field.  
By the central limit theorem, this change in field energy has an average value of zero and root mean square value proportional to $L^{d/2}$.
When the field energy savings are less than the interaction energy cost ($L^{d/2} < L^{d-1}$) the system remains ordered.
However, for $d \leq 2$, domain formation is favored, and the system is disordered for any nonzero disorder strength.  Smaller ratios of disorder strength to interaction strength correspond to larger characteristic domain sizes.

Returning to the EAHM, the correspondence with the RFIM shows that in the atomic limit and with $U=0$ the EAHM will have no CDW order for any nonzero disorder strength.
Moreover, hopping and onsite interactions both lower the energy cost of the boundaries.
Hopping makes the site occupancies continuous variables which can vary smoothly across domain walls.
Onsite interaction promotes single occupancy, and 
single occupancy corresponds in the RFIM to sites with zero spin.  Placed on a boundary, such sites lower the interaction-energy cost of the boundary.
We therefore do not expect true long-range CDW order in our disordered system.

Nonetheless, there is a crossover in the vicinity of $U=4V$, where there is a phase transition in the clean system.
For $4V<U$ ($V=0, 0.8$, and 1.6 in Fig.\ \ref{half_fill_Vdep}), onsite repulsion remains the dominant interaction, and the kinetic-energy driven ZBA found when $V=0$ persists.
When $4V>U$ ($V=2.4$ and 3.2 in Fig.\ \ref{half_fill_Vdep}), the nonlocal interaction dominates and the DOS appears to be dominated by atomic limit physics.

The case of $V=3.2$ is particularly simple to understand from an atomic limit perspective:
The DOS has a lower band and an upper band separated by a hard gap.
The location, width and shape of these bands are all consistent with the electrons forming a checkerboard pattern of alternating empty and doubly occupied sites.
The lower band corresponds to the removal of particles from doubly occupied sites.
The nearest neighbors of doubly occupied sites are all empty, so 
the DOS contribution of a single site with potential $\epsilon_i$ is $\epsilon_i + U - \mu$, with no dependence on $V$.
The site potentials $\epsilon_i$ are distributed between $-\Delta/2$ and $+\Delta/2$, creating a band in the DOS of width $\Delta$ centered on $U-\mu$.
The upper band corresponds to the the addition of particles to empty sites.
For an empty site there is no onsite energy cost to adding a particle, 
but the nearest neighbors of empty sites are all doubly occupied.
In this case, the DOS contribution of a single site with potential $\epsilon_i$ is $\epsilon_i + 8V - \mu$.
Again, there is a distribution of site potentials, resulting in a band of width $\Delta$ centered on $8V-\mu$.
In both cases the DOS slants downward toward the Fermi level.
This is because the ground state for a specific disorder configuration in a finite-size system will tend to have sites of especially low potential be doubly occupied and sites of especially high potential be empty.
This means that the lower band (coming from doubly occupied sites) will have somewhat more contributions from sites with low potentials and somewhat fewer contributions from sites with high potentials.
Note that this picture suggests that, although true CDW order is not expected, for these parameter values the size of the checkerboard domains is larger than our system size.

In this atomic-limit picture, for $V=2.4$ the separation between the centers of the two bands $8V-U$ is 11.2, slightly less than the disorder strength.  The two bands have therefore run together, but their slant downward toward the Fermi level is still apparent. 

\subsubsection{Spin and charge correlations}

\begin{figure}[htbp] 
   \centering
   \includegraphics[width=\columnwidth]{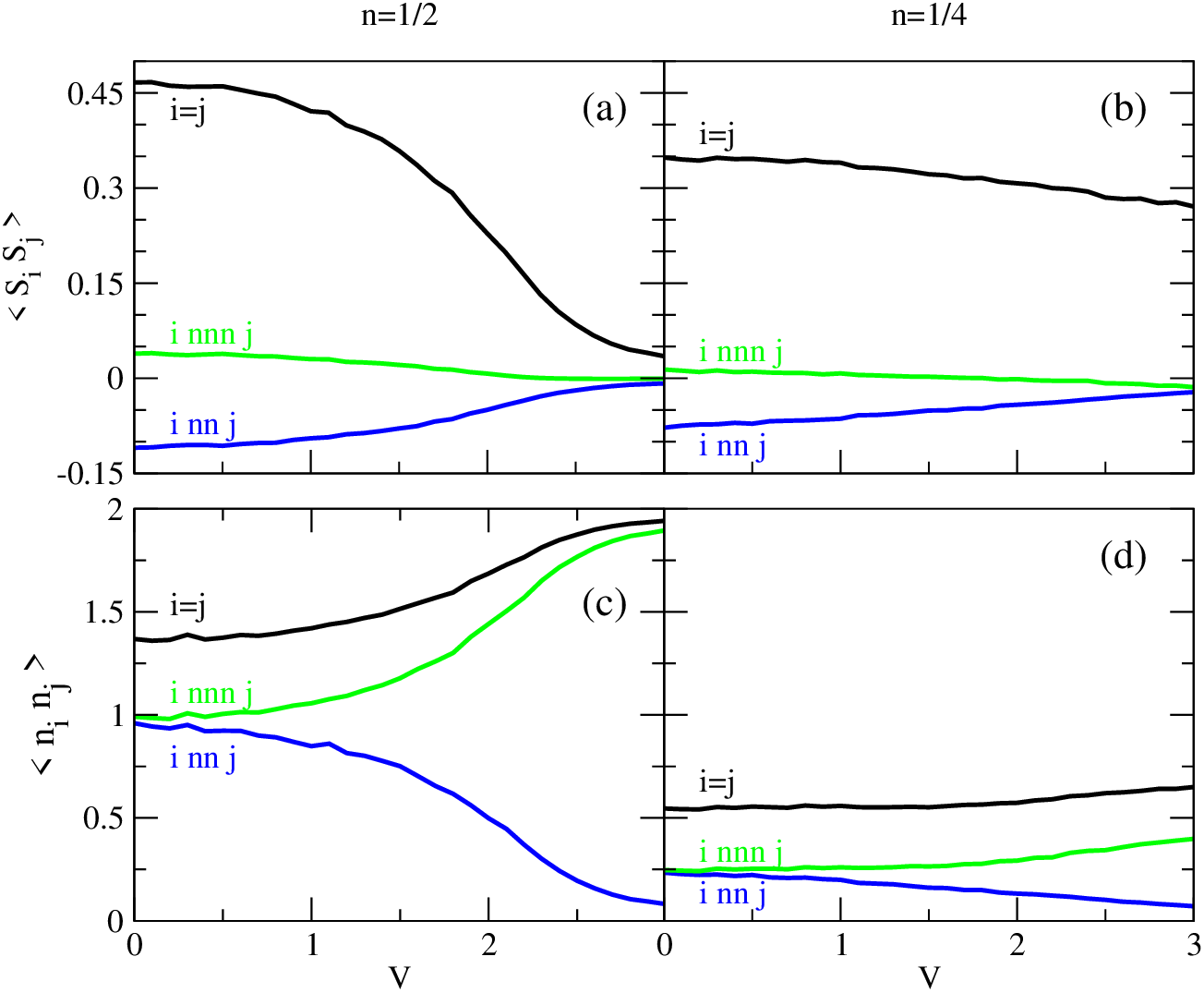} 
   \caption{Spin-spin correlation functions as a function of $V/t$ (a) at half filling and (b) at quarter filling.
Density-density correlation functions as a function of $V/t$ (c) at half filling and (d) at quarter filling.
$U=8t$ and $\Delta=12t$.}
   \label{correlations}
\end{figure}

To further highlight the crossover between $V\leq1.6$ and $V\geq2.4$, the spin and charge correlations are shown in Fig.\ \ref{correlations} (a) and (c).
For comparison, the spin correlations for a perfect singly occupied antiferromagnetic system are $+3/4$, $-1/4$ and $+1/4$ for on-site, nearest-neighbor, and next-nearest-neighbor respectively.  The corresponding charge correlations are one for all separations.
Meanwhile, the spin correlations for a checkerboard charge density wave with alternating doubly occupied and empty sites are all zero, and the charge correlations are 2, 0 and 2 for on-site, nearest-neighbor and next-nearest-neighbor.
Essentially these results are consistent with a crossover from a primarily singly-occupied and antiferromagnetically-correlated state at $V=0$ to a state with strong charge-density correlations at large $V$.
Note that the inflection point in these curves is at $U \sim 4V$.
The simultaneous washing out of the narrow, kinetic-energy-driven anomaly and the suppression of nonlocal spin correlations is consistent with the close association between these as highlighted in Ref.\ \onlinecite{Chen2011}.

\subsubsection{$U$ dependence at nonzero $V$}

\begin{figure}[htbp] 
   \centering
   \includegraphics[width=\columnwidth]{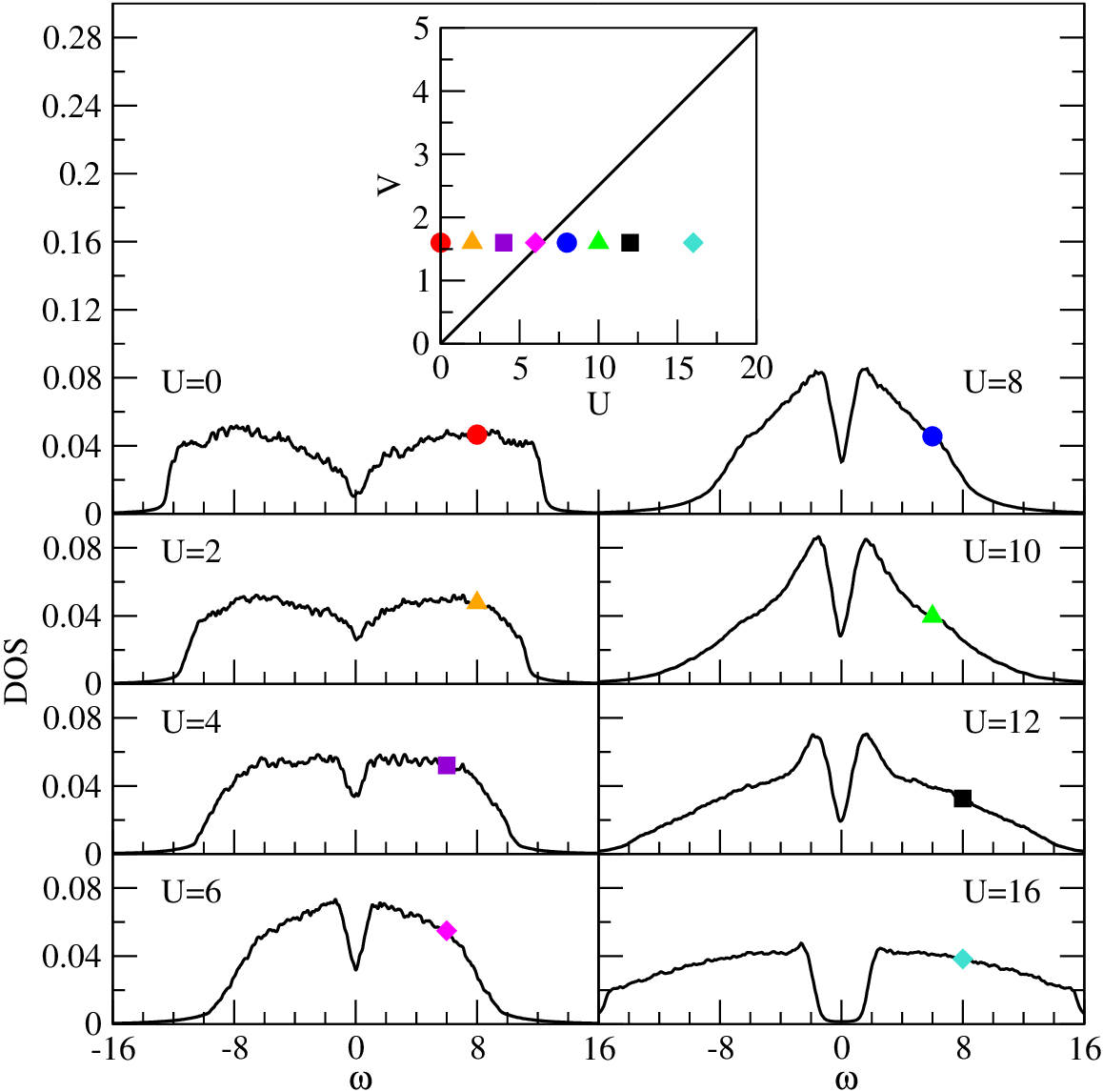} 
   \caption{Evolution of the DOS with $U/t$ at half filling for $\Delta/t=12$ and $V/t=1.6$.
   Inset shows the location of each parameter set in the clean phase diagram.}
   \label{Udep}
\end{figure}

Fig.\ \ref{Udep} shows a series of DOS results at fixed $V$ and increasing $U$.  The corresponding points on the phase diagram are indicated in the inset.
Here again we see a crossover from behavior consistent with the atomic-limit when $U<4V$ to features uniquely associated with the presence of hopping when $U>4V$.

For the two lowest $U$ values there is a very broad suppression of the DOS centered at the Fermi level.  
This is consistent with the picture discussed above of domains of checkerboard CDW order.
Again, the doubly occupied sites in these domains contribute to a plateau in the DOS of width $\Delta$ centered at $U-\mu$, while the singly occupied sites generate a corresponding plateau centered at $8V-\mu$.  
Both plateaus slant downwards towards the Fermi level due to the tendency for high potential sites to be empty and low potential sites to be full in the ground state. 
In the case of $U=0$, perfect charge ordering would result in a very narrow gap, $8V-U-\Delta=0.8$.
The presence of domain boundaries in a small number of disorder configurations would fill this in.
As $U$ is increased, the gap $8V-U-\Delta$ closes and in addition the energy cost of domain boundaries is lowered.  For both these reasons, the ZBA is weakened.

For $U=4$ and 6, the energy range of the DOS suppression is sharply reduced.  
This reduction in the width of the ZBA is not associated with hopping $t$, as it occurs in the atomic limit.\cite{Mulindwa2012}
An example is seen in Fig.\ \ref{tdep}(d):  The curve in Fig.\ \ref{tdep}(d) shows a ZBA which has roughly the same width as those in the $U=4$ and $U=6$ panels of Fig.\ \ref{Udep}.

At $U=8$, the onsite interaction is greater than $4V$ and the kinetic-energy-driven ZBA unique to strongly correlated systems emerges. 
This ZBA persists through $U=12$ with a consistent energy scale $\tilde t$, as discussed above.

Finally, at $U=16$, the Mott gap opens.  
When $V=0$, the Mott gap opens at $U \sim \Delta$.  
The addition of nonlocal interactions suppresses single occupancy, delaying the formation of the Mott gap and extending the range of the strong coupling ZBA.
Unlike the Mott gap, the strong coupling ZBA is not limited to half filling, and we now turn our attention to other dopings.

\subsection{Away from half filling}

\subsubsection{Doping dependence}

\begin{figure}[htbp] 
   \centering
   \includegraphics[width=\columnwidth]{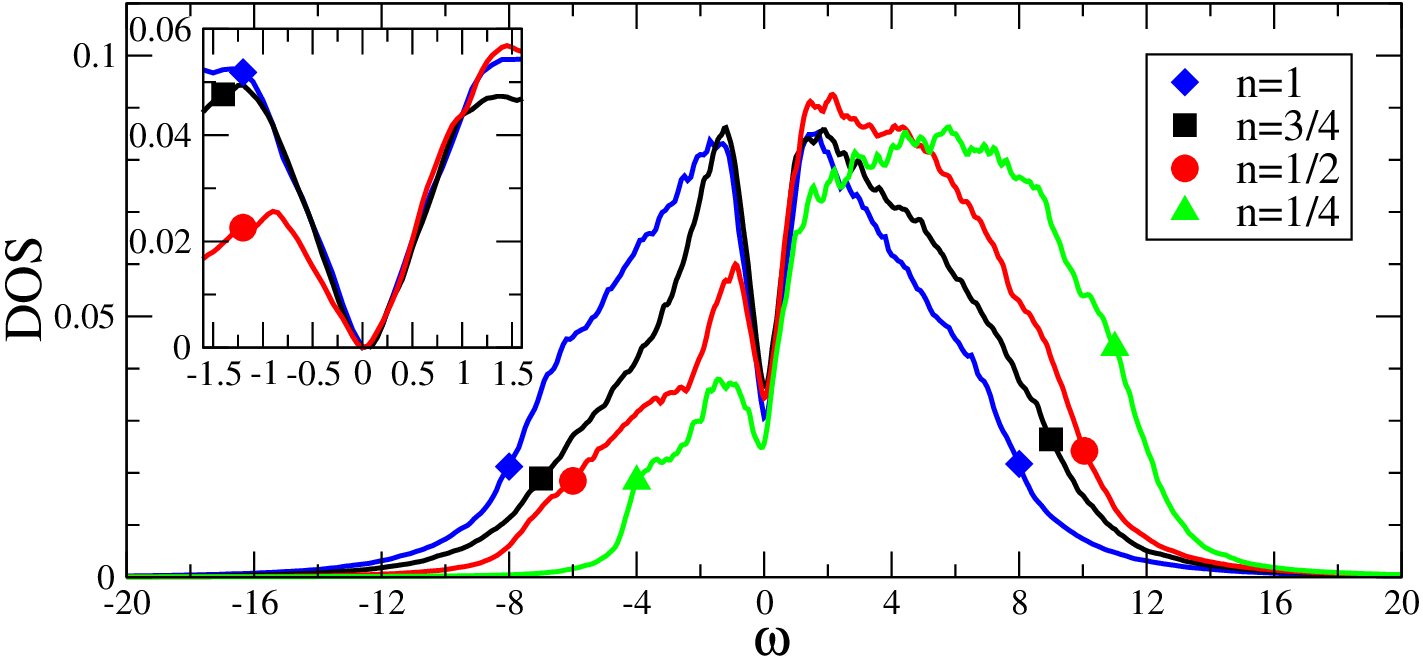} 
   \caption{Evolution of the DOS with filling for $\Delta/t=12$ , $U/t=8$ and $V/t=1.6$.}
   \label{ndep}
\end{figure}

Figure \ref{ndep} shows the dependence of the DOS on doping for $\Delta=12$, $U=8$ and $V=1.6$.  
The strong similarities between the curves is consistent with the picture that for $U>4V$ there is a kinetic-energy-driven ZBA proportional to $\tilde t$.  
With $t$ and $V$ both held constant, this strong coupling ZBA remains unchanged.
The deviation, especially below the Fermi level, of the $n=1/4$ curve is related to an important change in the atomic limit DOS:
For $V=0$ and in the atomic limit, above quarter filling there are doubly occupied sites and below there are none.
The Fermi level moves from inside the central plateau in Fig.\ \ref{tdep}(a) to its left edge.
As discussed above, similar energies of doubly occupied and singly occupied states on neighboring sites are central to the emergence of the kinetic-energy-driven ZBA.
While the presence of hopping smooths this transition, by $n=1/4$ doping the necessary conditions for a linear dependence on $\tilde t$ are absent.

\subsubsection{$V$ dependence at 1/4 filling}

\begin{figure}[htbp] 
   \centering
   \includegraphics[width=\columnwidth]{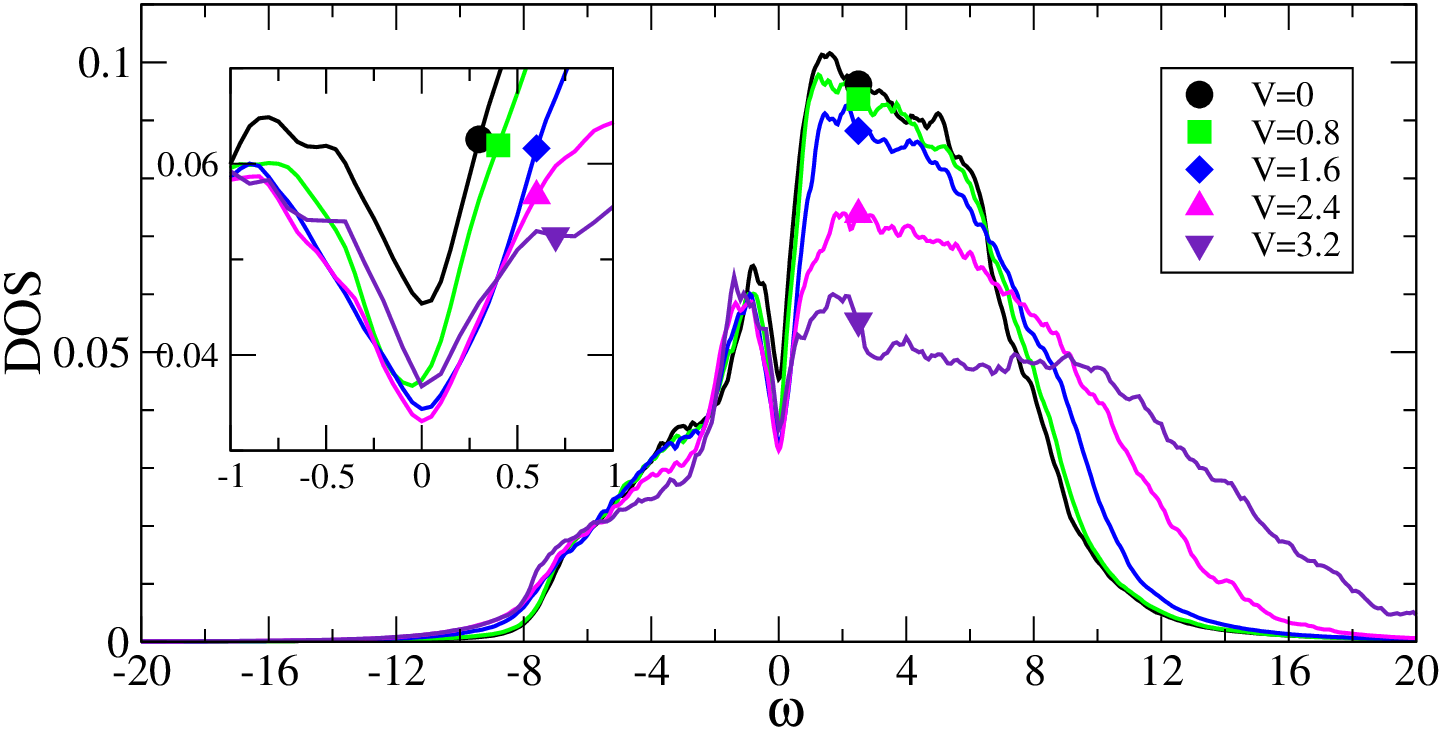} 
   \caption{Density of states versus frequency at quarter filling ($n=1/2$) for $\Delta=12t$, $U=8t$ and $V$ values as indicated. }
   \label{quarter_fill_Vdep}
\end{figure}

Figure \ref{quarter_fill_Vdep} shows the evolution of the DOS with $V$ at quarter filling.  Panels (c) and (d) of Fig.\ \ref{correlations} show the same mean-field rescaling of the DOS discussed for the half-filling case in Section \ref{small_V}.  
As at half filling, the rescaling works well for the $V=0.8$ case and also at very low energies for $V=1.6$.  

What is most striking about Fig.\ \ref{quarter_fill_Vdep}, however is the behavior at large $V$.
The spin and charge correlations at quarter filling, shown in Fig.\ \ref{correlations} (b) and (d), suggest a crossover similar to the one at half filling, although much weaker:
from AF correlated spins (and no CDW correlations) at low $V$ to CDW correlations (and no AF spin correlations) at large $V$.  
However, whereas at half filling (Fig.\ \ref{half_fill_Vdep}) these changes in correlations are matched by a suppression of the kinetic-energy-driven ZBA and the emergence of atomic-limit behavior, at quarter filling (Fig.\ \ref{quarter_fill_Vdep}) the form of the ZBA does not change when $4V>U$.
Lower filling reduces the competition between nearest-neighbor interactions and single occupancy.  
While $U$ reduces double occupancy and $V$ suppresses occupation of neighboring sites, both can be accommodated by a charge ordered state with alternating singly occupied and empty states.
Because of the importance of singly occupied sites to the formation of the kinetic-energy-driven ZBA, lower doping may allow it to persist to higher $V$ values.
It is noticeable, however, that persistence of behavior distinct from the atomic limit is not confined to the region around the Fermi level.
A number of other features seen at large $V$ are inconsistent with atomic-limit behavior:   
The onsite charge correlations increase with $V$, indicating the presence of double occupancy which is not expected in the atomic limit.
Also, the DOS remains essentially unchanged by $V$ below the Fermi level, with no sign of the broad DOS suppression expected in the atomic limit.
Although strong interactions reduce the importance of kinetic energy in a system, shifting it toward the atomic limit, lower electron concentration reduces this effect.



\section{Conclusion}
\label{sec-concl}

In conclusion, building on earlier work exploring the ZBA in the AHM and its unique strong-coupling features, here we ask how nonlocal interactions influence this anomaly.  
We find that at small values of $V$ and at low filling, there is no qualitative change in the anomaly, only a gradual renormalization of the hopping amplitude which sets its width.  
At larger values of $V$ and close to half filling, however, there is a crossover to DOS features which are independent of hopping.
As charge correlations grow, the energy scale of the kinetic-energy driven ZBA gives way to the higher energy scale of the nonlocal interaction.
The suggestion is that different strongly correlated materials with disorder may display very different behaviors depending on where their particular parameters place them in this phase space.
Moreover, it might be possible to observe such a crossover in a single material by, for example, applying pressure.
Increased pressure could be expected both to increase the hopping amplitude and also to reduce the nonlocal interaction through increased screening, driving the system from the atomic limit toward a regime in which the kinetic-energy driven zero-bias anomaly would appear.

\section*{Acknowledgments}
We acknowledge support by the National Science and Engineering Research Council (NSERC) of Canada.
This work was made possible by the facilities of the
Shared Hierarchical Academic Research Computing Network
(SHARCNET) and the High Performance Computing Virtual
Laboratory (HPCVL). 
H.-Y.C. is supported by NSC Grant No.
98-2112-M-003-009-MY3.

\end{document}